\documentclass[aps,prev,twocolumn,preprintnumbers,floatfix,nofootinbib]{revtex4-1}
\pdfoutput=1
\usepackage{amsmath}
\usepackage{amsfonts}
\usepackage{amssymb}
\usepackage{graphicx}
\usepackage{bm}
\usepackage{times}
\usepackage{hyperref}
\usepackage{slashed}
\usepackage{color}
\usepackage{color}

\begin{document}

\title{The Semi-Hooperon: Gamma-ray and anti-proton excesses in the Galactic Center}

\author{Giorgio Arcadi$^a$}
\email{arcadi@mpi-hd.mpg.de}

\author{Farinaldo S. Queiroz$^a$}
\email{queiroz@mpi-hd.mpg.de}

\author{Clarissa Siqueira$^{a,b}$}
\email{clarissa@mpi-hd.mpg.de}

\affiliation{$^a$Max-Planck-Institut f¨ur Kernphysik, Saupfercheckweg 1, 69117 Heidelberg, Germany}
\affiliation{$^b$Departamento  de  F\'{\i}sica,  Universidade  Federal  da  Para\'{\i}ba,
Caixa  Postal  5008,  58051-970,  Jo\~ao  Pessoa,  PB,  Brasil}

\begin{abstract}

A puzzling excess in gamma-rays at GeV energies has been observed in the center of our galaxy using Fermi-LAT data. Its origin is still unknown, but it is well fitted by Weakly Interacting Massive Particles (WIMPs) annihilations into quarks with a cross section around $10^{-26}{\rm  cm^3 s^{-1}}$ with masses of $20-50$~GeV,  scenario which is promptly revisited. An excess favoring similar WIMP properties has also been seen in anti-protons with AMS-02 data potentially coming from the Galactic Center as well. In this work, we explore the possibility of fitting these excesses in terms of semi-annihilating dark matter, dubbed as semi-Hooperon, with the process ${\rm WIMP\, WIMP \rightarrow WIMP\, X}$ being responsible for the gamma-ray excess, where X=h,Z. An interesting feature of semi-annihilations is the change in the relic density prediction compared to the standard case, and the possibility to alleviate stringent limits stemming from direct detection searches. Moreover, we discuss which models might give rise to a successful semi-Hooperon setup in the context of $\mathcal{Z}_3$,$\mathcal{Z}_4$ and extra ``dark'' gauge symmetries.
%dark $SU(3)$ symmetries.

\end{abstract}

%\pacs{95.35.+d, 14.60.Pq, 98.80.Cq, 12.60.Fr}

\maketitle

\section{Introduction}

Dark matter is one of the pillars of the Standard Cosmological Model, but its nature is unknown, despite the compelling evidence at different time scales of the universe evolution and distance scales of the universe today. Since the Standard Model of particle physics has no particle able to account for the presence of dark matter in our universe, theories beyond the Standard Model are needed. Therefore, the nature of dark matter is one of the most important open problems in modern cosmology and particle physics, as can be seen by the enormous theoretical and experimental effort being put forth toward its identification. 

One of the most compelling candidates are the Weakly Interacting Massive Particles (WIMPs) \cite{Jungman:1995df, Bergstrom:2000pn, Bertone:2004pz,Arcadi:2017kky}. A signal in indirect dark matter detection represents a major step towards unveiling the nature
of dark matter, and from this perspective, gamma-rays play a key role. WIMPs are, indeed, expected to shine in gamma-rays and these enable us to trace back the source, serving as a good handle for background discrimination \cite{Bringmann:2012ez}.

Albeit, it is quite challenging to disentangle a potential DM signal from the large astrophysical foreground and backgrounds that dominate the measured gamma-ray flux. Thus, two searches for a DM signal stand out: signal maximization or background reduction. In the former, targets such as the Galactic Center (GC) are used, whereas in the latter Dwarf Spheroidal Galaxies (dSphs) become ideal hosts \cite{GeringerSameth:2011iw,Conrad:2014tla,Geringer-Sameth:2014qqa,Bonnivard:2015tta,Mambrini:2015sia,Baring:2015sza,Geringer-Sameth:2015lua,Queiroz:2016zwd}.

In particular, the GC is expected to be the brightest source in gamma-rays \cite{Abramowski:2011hc, Hooper:2012sr}, but our current knowledge of the astrophysical foreground and
background is subject to large uncertainties. For this reason is not surprising to spot gamma-ray excesses towards the inner region of our galaxy.  Anyways,
several groups reported the observation of a residual gamma-ray emission using the publicly available gamma-ray data from the Fermi-LAT satellite \cite{Goodenough:2009gk,
Vitale:2009hr,
Hooper:2010mq,Hooper:2011ti,Abazajian:2012pn, Gordon:2013vta,Macias:2013vya,
Abazajian:2014fta,Daylan:2014rsa,Zhou:2014lva,Carlson:2014cwa}. The excess is peaked around $1-3$~GeV, and has spectral and morphological features similar to those expected from DM annihilations.

Excitingly, the excess has been confirmed by the Fermi-LAT collaboration \cite{TheFermi-LAT:2015kwa} to enliven the signal, and its seems to be extended up to higher latitudes as expected from dark matter annihilation \cite{Hooper:2013rwa,Daylan:2014rsa}. Past studies have been done to determine the best-fit region for the gamma-ray excess in terms of dark matter annihilations \cite{Huang:2013pda,Agrawal:2014oha,Calore:2014nla}, motivating model building endeavors in terms of WIMP annihilations.

Moreover, more recently an excess in anti-protons has been observed in the AMS-02 data \cite{Cuoco:2016eej,Cui:2016ppb}. The source of this excess is unknown but it is quite plausible to come from the GC. This excess seems similar to one firstly observed in the PAMELA data \cite{Hooper:2014ysa}. Anyways, in \cite{Cuoco:2016eej} the authors went beyond the usual benchmark propagation models MIN/MED/MAD scenarios that are motivated by the Boron over Carbon ratio \cite{Donato:2003xg}, and performed a global fit to the AMS-02 anti-proton data to claim an over $4\sigma$ signal from dark matter annihilation. As described in \cite{Cuoco:2017rxb}, the WIMPs properties favored by this signal are similar to those from the GeV excess in gamma-rays. Therefore, in what follows, we will assume that the processes capable of explaining the GeV gamma-ray excess are also providing a good fit to the anti-proton excess. For this reason, whenever we refer to the GeV excess in gamma-rays bear in mind that there is also a similar excess in anti-protons.

However recent null results from direct dark matter detection experiments and colliders have shrunk quite a bit the number of models capable of accounting for the gamma-ray excess \cite{Alvares:2012qv,Alves:2014yha,Hooper:2014fda,Agrawal:2014una,Queiroz:2014yna,Berlin:2014pya,Berlin:2014tja,Hooper:2014fda,Freytsis:2014sua,Cline:2014dwa,Boehm:2014bia,Cheung:2014lqa,Alves:2015mua,Berlin:2015wwa,Alves:2015pea,Modak:2015uda,Kaplinghat:2015gha,Gherghetta:2015ysa,Duerr:2015bea,Elor:2015tva,Kim:2015fpa,Kopp:2015bfa,Alves:2016cqf,Karwin:2016tsw,Escudero:2016kpw,Choquette:2016xsw,DuttaBanik:2016jzv,Boddy:2016hbp,DuttaBanik:2016jzv,Kim:2016csm,Li:2016uph,Jia:2017kjw,Nomura:2017emk,Okada:2014usa,Batell:2017rol,Eiteneuer:2017hoh} in agreement with existing bounds. Therefore, it is worthwhile to investigate different setups. 

That said, after briefly reviewing the status of dark matter annihilations as an explanation to the GC excess, we discuss the possibility of explaining the gamma-ray excess in the GC via semi-annihilations~\cite{DEramo:2010keq,Belanger:2012zr,DEramo:2012fou,Belanger:2012vp,Ko:2014loa,Aoki:2014cja,Belanger:2014bga,Cai:2015tam,Cai:2015zza,Cai:2016hne}. By semi-annihilations one typically refers to processes of the type $DM\, DM \rightarrow DM\, X$ involving three DM particles, two in the initial state and one in the final state, and a state $X$ which can be either an SM state or an additional (unstable) exotic state. It can be noted that semi-annihilations occur in scenarios where the DM is made stable via symmetries larger than $\mathcal{Z}_2$, as for example $\mathcal{Z}_3$. 

In most of the WIMP scenarios, direct and indirect detection observables are tightly connected. This feature is particularly evident in the so-called simplified models~\cite{Alwall:2008ag,DiFranzo:2013vra,Balazs:2014jla,Berlin:2014tja,Buckley:2014fba,Abdallah:2014hon,Baek:2015lna,Kahlhoefer:2015bea,Bell:2015rdw,Choudhury:2015lha,Abdallah:2015ter,Liew:2016oon,Sandick:2016zut,Bell:2016uhg,DeSimone:2016fbz,Boveia:2016mrp,Brennan:2016xjh}, where DM annihilations and the WIMP-nucleon scattering cross section are governed by the same mediator and couplings. Semi-annihilation that can have a strong impact on the DM relic density, on the other hand, are not necessarily connected to the WIMP-nucleon scattering rate. Therefore, sizeable semi-annihilations without conflicting direct detection limits are feasible. For this reason, semi-annihilations represent a plausible mechanism to explain the GeV gamma-ray excess in agreement with existing data. 

As we shall see further, indirect detection limits are still strong as in the case of annihilating dark matter, since a good fit to the GeV excess through semi-annihilation requires a relatively large semi-annihilation cross-section which can be probed with indirect detection experiments. We emphasize that the crucial aspect of models with dominant semi-annihilations is the possibility to weaken the restrictive limits from direct detection experiments which typically force the existing models to live in regions of parameter space where resonance effects are dominant. \\

In summary, in this work we will assess the capability of two specific semi-annihilation processes, $DM DM \rightarrow DM Z$ and $DM DM \rightarrow DM h$, with $Z$ and $h$ being the SM $Z$ and $h$ bosons, of yielding a good fit of the GeV excess. Pursuing a model-independent approach we will express our results in terms of best-fit DM mass and semi-annihilation cross-section, in the considered final states, without referring to any specific Particle Physics construction. In the second part of the paper, we will provide a brief overview of some possible models, involving scalar, fermionic and vectorial DM, potentially capable of reproducing the GeV excess through semi-annihilations and featuring, at the same time, viable DM relic density compatible with observational limits.

\section{The Gamma-Ray Excess}

Interestingly, since the first observation \cite{Goodenough:2009gk}, independent analyses of the Fermi-LAT data toward the Galactic Center have spotted an excess in gamma-rays peaked around $3$~GeV \cite{Vitale:2009hr,Hooper:2010mq,Abramowski:2011hc,Abazajian:2012pn,Bringmann:2012ez,Hooper:2012sr,Gordon:2013vta,Macias:2013vya,Abazajian:2014fta,Daylan:2014rsa,Zhou:2014lva}, which is also known as the GeV excess. Though the excess is statistically significant, it relies on a diffuse background model which provides a reasonable fit to the diffuse emission over the whole sky, but it is not clear that it is appropriate for the GC. This systematic uncertainty on the background
dominates the statistical one and is difficult to quantify. A first attempt at accounting for
systematics was made in~\cite{Calore:2014nla} through the study of a large number of galactic diffuse emission models, where the excess persisted. Subsequent studies performed by Fermi-LAT collaboration reached the same conclusion \cite{TheFermi-LAT:2015kwa}. In particular, in \cite{TheFermi-LAT:2017vmf} it has been concluded that is most likely comprised of two components, where one of them is from dark matter.
Therefore, we will assume the signal is robust and due to dark matter annihilations, commonly known as Hooperons\cite{Alves:2014yha}.

\section{The Anti-Proton Excess}

Several studies have been conducted regarding the anti-proton flux originated from dark matter annihilation \cite{Chardonnet:1996ca,Bergstrom:1999jc,Boehm:2009vn,Perelstein:2010gt,Garny:2011cj,Evoli:2011id,Fornengo:2013xda}. Throughout the years no excess has been observed except in \cite{Hooper:2014ysa} where an excess has been spotted using PAMELA data. Analysis favored a dark matter annihilation cross section which has now been ruled by the non-observation of gamma-rays in dSphs by Fermi-LAT collaboration \cite{Fermi-LAT:2016uux}. Recently, however, two independent studies, using up-to-date data from AMS-02 claimed the observation of a significant excess of anti-proton production over the background expectations \cite{Cuoco:2016eej,Cui:2016ppb}. 

This excess regardless its origins clearly shows the importance of data taking and further extending AMS-02 lifetime. The main differences between these two studies and previous ones where no excess was observed are the use of new data and a global likelihood analysis, where both background and signal are simultaneously fitted, without assuming a fixed propagation model. Keep in mind that this signal is subject to several systematic uncertainties and rely on the non-existence of secondary production of cosmic-rays, which could significantly abate the signal \cite{Feng:2016loc,Cholis:2017qlb,Winkler:2017xor,Yuan:2017ozr,Niu:2017qfv}.

The recent analysis presented in \cite{Cuoco:2017rxb} clearly shows that the excess in anti-proton observed in the AMS-02 data strongly overlaps with the one seen by Fermi-LAT in the Galactic Center. Therefore, instead of doing data fitting for independent data sets and repeat what has been done in \cite{Cuoco:2017rxb}, we will concentrate on one of them (gamma-ray data), and assume throughout that the best-fit regions derived for the annihilation and semi-annihilation cases coincide with those from anti-proton.

Moreover, we emphasize as suggested in the title that this source of anti-proton should be local with a large J-factor, otherwise the best-fit region in the plane {\it annihilation cross section vs dark matter mass} would be moved upwards being in direct conflict with the non-observation of gamma-rays in the direction of dwarf galaxies as seen by Fermi-LAT.

\section{Fitting the Data} \label{model}

In order to find the goodness of fit to the GeV excess, we need to compute the differential photon flux in the direction of the GC which is written as,

%The production of photons is dominated by production of SM particles which
%subsequently produce photons through decays, or to a lesser extent bremsstrahlung. The differential flux of such photons from a %given direction is given by
%
\begin{equation}
\frac{d\Phi^\gamma}{dE^\gamma}=\sum_f \frac{1}{8\pi}\frac{ \langle\ \sigma v_f \rangle }{M_{DM}^2}\frac{dN^\gamma_f}{dE^\gamma} \times \bar{J}(\Delta\Omega),
\label{Eq:flux}
\end{equation}
where the sum runs over all the annihilation final states considered in the analysis of the signal. $\langle \sigma v _f \rangle$ is the annihilation cross section today into a final state $f$ with $v \sim 10^{-3} c$, $M_{DM}$ is the DM mass, the $dN^\gamma_f/dE^\gamma$ is the DM prompt gamma-ray spectrum for each $f$ final state, and $\bar{J}(\Delta\Omega)$ is the J-factor, which is given by:
\begin{equation}
\bar{J}(\Delta\Omega)= \frac{1}{\Delta \Omega}\int_{\Delta\Omega}\int_{l.o.s.}\rho^2(r(s,b,l))dsd\Omega,
\end{equation}
where $r(s,b,l)=(r_{\odot}^2+s^2-2\, r_\odot s \,\cos b \cos l)^{1/2}$, and $r_\odot$  the distance between the GC and the Sun ($8.5 \, {\rm kpc}$), with our region of interest (ROI) for the galactic latitude, $2^\circ \leq |b| \leq 20^\circ$, and for the galactic longitude, $0^\circ\leq |l| \leq 20^\circ$.
This astrophysical J-factor was computed considering a generalized Navarro-Frenk-White (gNFW) profile \cite{Navarro:1996gj,Navarro:1995iw}, thus, the expression for local DM density is given by
\begin{equation}
\rho(r)= \rho_0 \frac{(r/r_s)^{-\gamma}}{(1+r/r_s)^{3-\gamma}},
\end{equation}
where the $\rho_0$ is the scale density calculated by fixing the local DM density at the Sun on $\rho_{\odot}= 0.4 \,{\rm GeV/cm^3}$, the scale radius $r_s=20\,{\rm  kpc}$ and $\gamma=1.2$.

With all these ingredients we can compute the goodness of fit to the data following the receipt described in \cite{Calore:2014nla} where the 
$\chi^2$ is defined as,
\begin{equation}
\chi^2=\sum_{i,j}\left[ \left(\frac{d\Phi^\gamma_i}{dE^\gamma} - \frac{dF_i}{dE^\gamma} \right) (\Sigma^{-1})_{ij}  \left(\frac{d\Phi_j^\gamma}{dE^\gamma}_j - \frac{dF_j}{dE^\gamma}\right) \right],
\label{Eq:chi2}
\end{equation}where $d\Phi^\gamma_i/dE^\gamma$ is the flux predicted and $dF_i/dE^\gamma$ is the observed flux for the i-th energy bin and $\Sigma_{ij}$ is the covariance matrix which contains the statistical and correlated systematical errors, for the 24 energies bins provided in \cite{CCW}. Thus, in total, we have 22 degrees of freedom, 24 energies bins minus two degrees stemming from the annihilation cross section and dark matter masses. In what follows we describe this procedure in terms of dark matter annihilations.

%In the following, we will review the status of annihilation channels fitting GCE and, in the next, we will present our results for the case of semi-Hooperon.

\subsection{Annihilations}

Dark matter annihilations yield a gamma-ray flux as predicted by Eq.\eqref{Eq:chi2}. After fixing the J-factor and choosing an annihilation final state, which dictates the energy spectrum computed with PPPC4DMID \cite{Cirelli:2010xx}, the two unknown are the dark matter annihilation cross section and mass. Thus, one can now plug Eq.\eqref{Eq:flux} into Eq.\eqref{Eq:chi2}, and find the best-fit regions in the ${\it \langle\sigma v\rangle\, vs\, \rm M_{DM} }$ plane.

In the Table \ref{tableI}, we show the best-fit regions for several annihilation final states, including the $1\sigma$ error for the mass and annihilation cross section, the minimal $\chi^2$ and the p-value. We can see on Table \ref{tableI} that there is a mild preference for DM to annihilation into quarks ($\bar{b}b$ and $\bar{q}q$) and Higgs particles. For the $\bar{\mu}\mu$ channel we are considering just the prompt contribution, however, if we include the ICS (depending on the Galactic diffuse emission model) we can find a better fit to data \cite{Calore:2014nla}.  The $\bar{\tau}\tau$ channel yields a slightly worse fit to data, whereas the $ZZ$, $WW$ and $\bar{t}t$ channels a significantly larger $\chi^2$. However, it should be noted that the characterization of resolved point sources might significantly affect the spectrum of the residual emission \cite{TheFermi-LAT:2017vmf}, so the difference in the $\chi^2$ arising between various annihilation final states should be taken with a grain of salt. In summary, arguably all these final states offer a good fit the GeV excess. 

\begin{table}[!h]
\centering
\begin{tabular}{|c|c|c|c|c|}
	\hline
	Channel & Mass & $\langle \sigma v \rangle$ & $\chi^2_{min}$ & p-value \\
     & (GeV) &  ($10^{-26}cm^3 s^{-1}$) & & \\
	\hline
    $\bar{b}b$ & 48.70$^{+7.96}_{-6.14}$ & 1.79 $\pm$ 0.35 & 24.46 & 0.3236 \\
	\hline
    $\bar{q}q$ & 23.91$^{+3.47}_{-3.21}$ & 0.83 $\pm$ 0.16 & 26.93 & 0.2139 \\
    \hline
    $hh$ & 125.70$^{+4.10}_{-0.00}$ & 5.37 $\pm$ 1.03 & 29.24 & 0.1381 \\
    \hline
    $\bar{\tau}\tau$ & 9.96$^{+1.81}_{-1.66}$ & 3.39 $\pm$ 0.79 & 34.28 & 0.0460 \\
    \hline 
    $ZZ$ & 91.20$^{+2.94}_{-0.00}$ & 4.11 $\pm$ 0.84 & 40.36 & 0.0098 \\
    \hline
    $WW$ & 80.40$^{+2.72}_{-0.00}$ & 3.38 $\pm$ 0.72 & 42.96 & 0.0048 \\
    \hline
    $\bar{\mu}\mu$ & 5.03$^{+0.80}_{-0.06}$ & 1.82$^{+0.47}_{-0.38}$ & 46.78 & 0.0016 \\
	\hline
    $\bar{t}t$ & 173.30$^{+4.00}_{-0.00}$ & 5.25 $\pm$ 1.27 & 53.05 & 0.0002 \\
    \hline 
\end{tabular}
\caption{Best-Fit regions for the GCE for different channels including mass, cross section, $\chi^2$ and p-value.}
\label{tableI}
\end{table}

\begin{figure}[!h]
\centering
\includegraphics[width=\columnwidth]{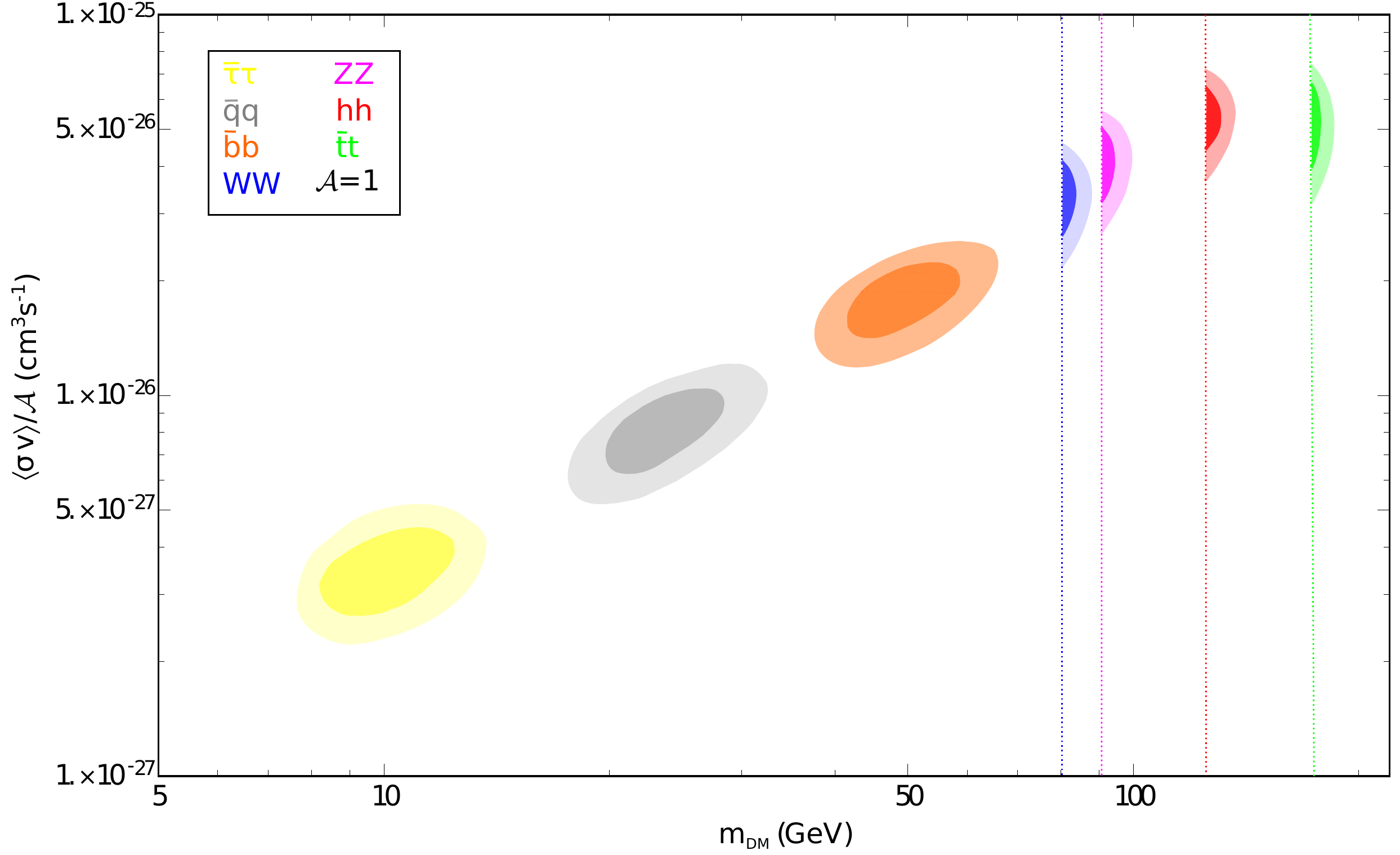}
\caption{The viable $1\sigma$ and $2\sigma$ regions for each channel taking $\mathcal{A}=1$, where A encompasses the uncertainty in the J-factor. We assumed that the DM particle annihilates with a 100\% branching ratio into a given final state at a time throughout.}
\label{Graph1}
\end{figure}

In the Fig. \ref{Graph1} we summarize all this information. We show the $1\sigma$ and $2\sigma$ regions for each channel in the $\langle \sigma v \rangle$ \textit{versus} $M_{DM}$ plane able to fit the GeV excess. Similarly to~\cite{Calore:2014nla} we have parametrized the uncertainties related to the choice of the DM halo model through a multiplicative parameter $\mathcal{A}$. The results reported in Fig.~\ref{Graph1} have been obtained by setting $\mathcal{A}=1$. 

Setting aside the shift in the best-fit regions caused by the difference in the annihilation final state, an important message we take from the GeV excess is the favored annihilation cross section about $10^{-26} cm^3 s^{-1}$ which is close to standard annihilation cross section that yields the right DM relic density.

Typically, in dark matter models the same couplings govern the relic density and the scattering cross section   \cite{Profumo:2013sca,Alves:2013tqa,Queiroz:2014pra,Allanach:2015gkd,Arcadi:2017atc}. Therefore, null results from direct detection experiment can severely constrain the annihilation rate and consequently a possible explanation to the astrophysical signals under discussion.  This scenario where annihilation is the main or only contribution to the relic density and the GeV excess also known as Hooperon is tightly restricted by direct detection experiment. For this reason it is worthwhile to go beyond the Hooperon regime, and discuss the impact of semi-annihilations.

\subsection{Semi-Annihilations}
For now, we will consider the case of semi-annihilation channels. Let's assume that  the DM could self-annihilate in the channel DM X, where $X$ can be a Z boson or the standard Higgs particle $h$. Considering that the DM can annihilate $100\%$ in each channel, we compute the gamma-ray flux using the Pythia routines \cite{Sjostrand:2014zea}, since PPPC4DMID does not include semi-annihilations \cite{Cirelli:2010xx}. The exact derivation of the best-fit regions for semi-annihilating dark matter is one of the main features of this work, since it has not been derived before. 

In the Table \ref{tableII} we report the best fit for each channel, including the preferable mass, and the thermally averaged cross section with $1\sigma$ error, and $\chi^2$ and the p-value for each case, where the p-value has been calculated using a Gaussian distribution. In both cases we have the mass threshold for kinematic reasons. Moreover comparing our results with the results for annihilation channels showed in Table \ref{tableI} we found the same $\chi^2$ and, consequently, p-value, but for larger cross sections. One can easily conclude that semi-annihilations into $X\,h$ lead to a better fit than those into $X\,Z$. Indeed, semi-annihilation into Z, is sufficiently poor to yield a p-value smaller than the standard value $0.05$.
\begin{table}[!h]
\centering
\begin{tabular}{|c|c|c|c|c|}
	\hline
	Channel & Mass & $\langle \sigma v \rangle$ & $\chi^2_{min}$ & p-value \\
     & (GeV) &  ($10^{-26}cm^3 s^{-1}$) & & \\
    \hline
    DM $h$ & 125.7$^{+6.20}_{-0.00}$ & 10.7 $\pm$ 2.10 & 29.24 & 0.1381 \\
    \hline
    DM $Z$ & 91.2$^{+4.82}_{-0.00}$ & 8.22 $\pm$ 1.73 & 40.36 & 0.0098 \\
    \hline
\end{tabular}
\caption{Best-Fit regions for the GCE for semi-annihilation channels including mass, cross section, $\chi^2$ and p-value with $1\sigma$ error.}
\label{tableII}
\end{table}

\begin{figure}[!h]
\centering
\includegraphics[width=\columnwidth]{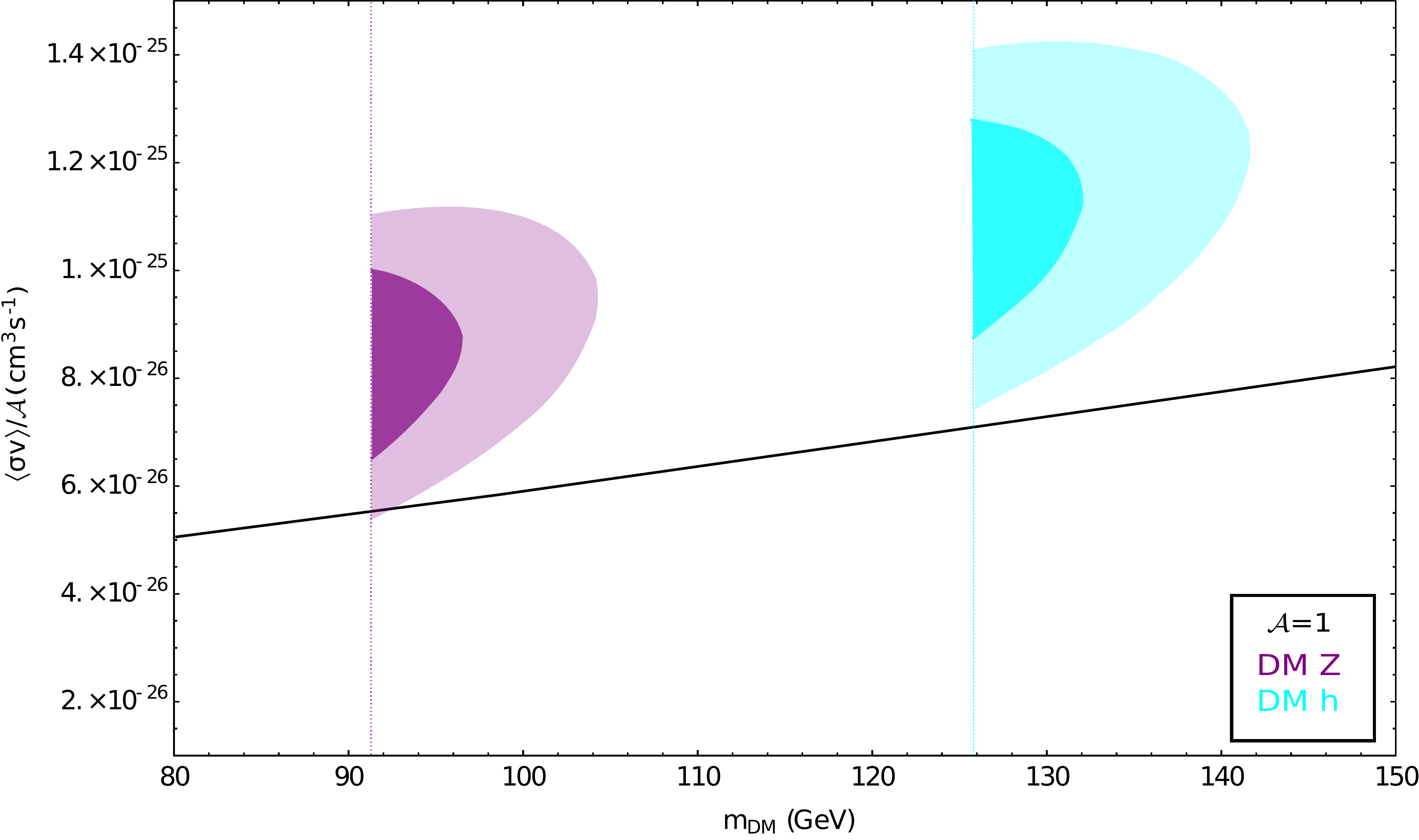}
\caption{The viable and excluded  $1\sigma$ and $2\sigma$ regions for for h and Z channel taking $\mathcal{A}=1$, respectively.}
\label{Graph2}
\end{figure}

\begin{figure}[!h]
\centering
\includegraphics[width=\columnwidth]{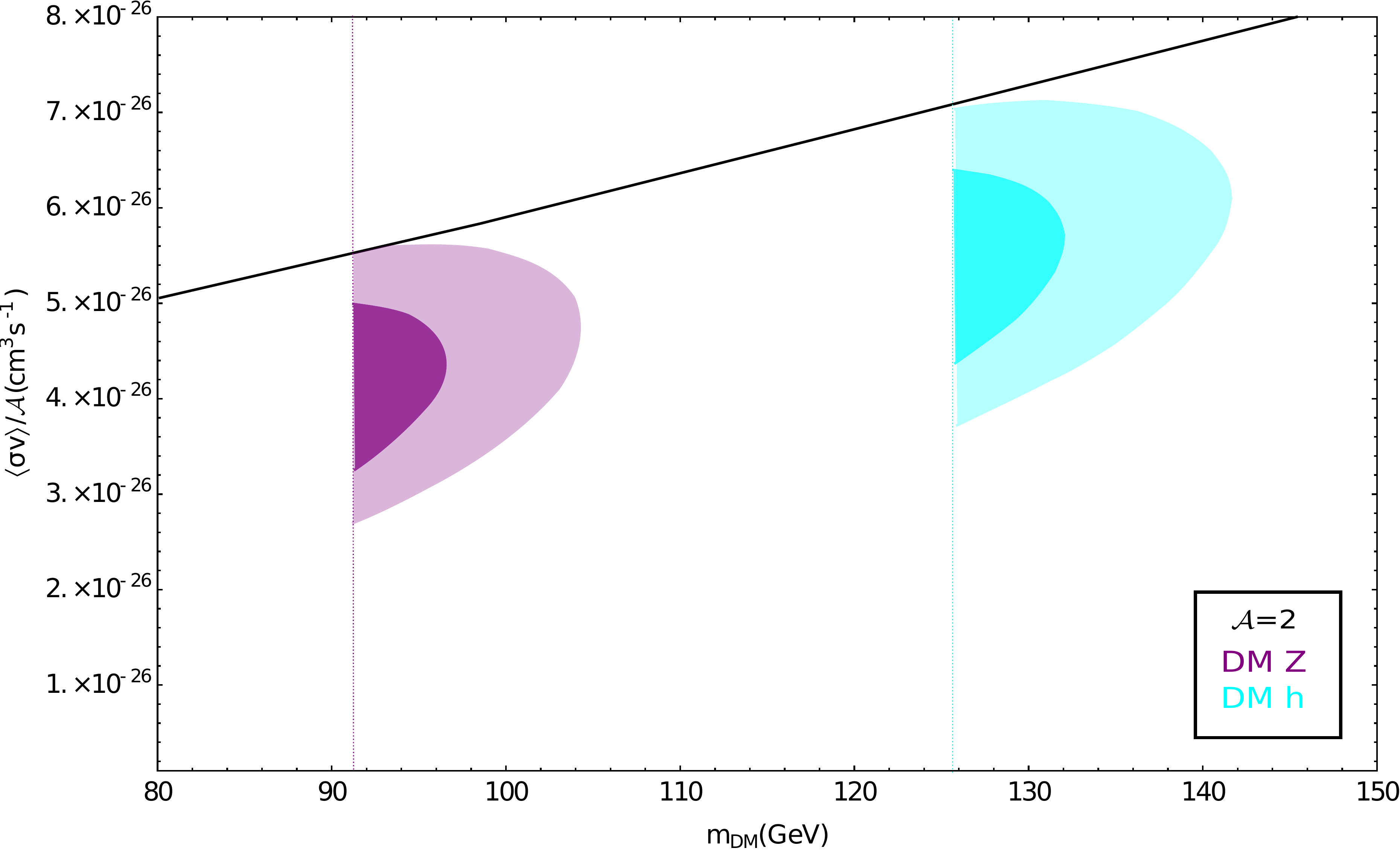}
\caption{The viable and excluded $1\sigma$ and $2\sigma$ regions for h and Z channels taking $\mathcal{A}=2$, respectively.}
\label{Graph3}
\end{figure}

Anyways, in the Figs. \ref{Graph2}-\ref{Graph3} we show the regions on the plane $\langle\sigma v \rangle$ -$M_{DM}$ with $1\sigma$ and $2\sigma$ error for the channels DM h, which provide the best-fit to the GeV excess. In these plots were adopted $\mathcal{A}=1-2$. We also include the upper limits over the annihilation cross section imposed by 6 years of Fermi-LAT Data telescope derived from the stack-analysis of gamma-ray observation from Dwarf Spheroidal Galaxies (dSphs) \cite{Ackermann:2015zua}. In the derivation of these limits we used the public likelihood functions provided by the Fermi-LAT collaboration. 

Notice that for $\mathcal{A}=1$, regardless of how good the fit to data is, gamma-ray observations from dphs basically exclude both semi-annihilation setup as possible explanations to the GeV excess.
On the other hand, if the J-factor in the galactic is larger by a factor of two, the best-fit regions are moved downwards by a factor two to yield the same $\chi^2$, as can be seen in Fig.\ref{Graph3}. Since the uncertainties over the DM density in the GC are rather large, the value $\mathcal{A}=2$ is plausible. The importance of adopting $\mathcal{A}=2$ is visible in Fig.\ref{Graph3} where now the entire best-fit regions for the semi-hooperon are consistent with dphs observations \footnote{Notice that this shift in the best-fit regions by taking $\mathcal{A}=2$ would also occur for self-annihilations. }

As we discussed previously, simplified models of self-annihilating dark matter suffer from stringent limits from direct detection experiments. For this reason, it is important to investigate models beyond this setup and that has motivated us to precisely derive the best-fit regions for semi-annihilation channels. Now we discuss which models would feature these processes, while potentially being capable of explaining the excesses in gamma-rays and anti-protons.

\subsection{Models for Semi-annihilations}

\noindent
We will give an overview of some models potentially capable of reproducing the excesses in gamma-rays and anti-proton in the semi-hooperon setup. In light of the results of the previous section we will focus just on the $DM\, DM \rightarrow DM\,h$ process.

Model building of semi-annihilating DM models is not trivial since it requires considering interactions involving an odd number of DM particles while ensuring the right relic dark matter density.

As can be easily argued the minimal option would be to consider the introduction of a $\mathcal{Z}_3$ discrete symmetry, allowing (complex) scalar or vectorial DM candidates to interact with the SM Higgs through four field renormalizable interaction terms. Semi-annihilating DM could be also achieved in presence of different symmetry groups provided that the DM belongs to a non-trivial ``hidden'' sector, i.e. composed by different states (not necessarily all stable) suitably charged under the symmetry group responsible of the stability of the DM. Semi-annihilation can, in particular, easily occur in models featuring multi-component DM. An extensive discussion of semi-annihilation models has been done, at the level of Effective Field Theory (EFT), in~\cite{Cai:2016hne}. We will rather discuss below some more specific examples.
Before doing so, we emphasize that while such discrete symmetries might plausibly  arise from an Abelian gauge group \cite{Patra:2016ofq,Klasen:2016qux,Alves:2016fqe}, the use of global symmetries to spontaneously generate a discrete symmetry is disfavored \cite{Mambrini:2015sia,Alves:2016bib}.

\subsubsection{$\mathcal{Z}_3$ singlet scalar DM}

The simplest model featuring semi-annihilating consists in the extension of the SM Higgs sector by a SM singlet complex scalar field $S$ transforming under a discrete $\mathcal{Z}_3$ so that one can write the following scalar potential:
\begin{align}
\label{eq:Vsinglet}
& V=\mu_H |H|^2+\lambda_H |H|^4+\mu_S^2 |S|^2+\lambda_S |S|^4\nonumber\\
& +\lambda_{SH} |S|^2 |H|^2+\frac{\mu_S^{'}}{2}\left(S^3+S^{\dagger\,3}\right)
\end{align}
invariant under the transformation $H \rightarrow H,S\rightarrow {e}^{\frac{i 2\pi}{3}}S$.
The DM relic density is, in general, determined both by normal pair annihilations, whose rate depend on $\lambda_{SH}$ (this coupling is also responsible of spin independent interactions with the nuclei), and by the semi-annihilation $SS \rightarrow S^{\dagger}h$, determined instead by the coupling $\mu_S^{'}$. The latter becomes the dominant process for DM relic density provided that $\mu_S^{'} \gg \lambda_{SH}$. 

Notice that the annihilation processes into fermions occur via the Higgs portal, i.e. proportional to $\lambda_{HS}$, whereas semi-annihilation that appears through the t-channel exchange of the scalar $S$ is proportional to $\lambda_{SH}^2 \mu_S^{' 2}/m_S^6$. Therefore, once can enhance semi-annihilations by increasing $\lambda_{SH}$. Indeed, defining $\alpha$ the fraction of semi-annihilation,

\begin{equation}
\alpha = \frac{1}{2} \frac{\sigma v (S S \rightarrow S^{\dagger} h)}{ \sigma v (S S^{\dagger} \rightarrow X X) + 1/2\sigma v (S S \rightarrow S^{\dagger} h)},
\label{alphaEq}
\end{equation}one can easily conclude that,

\begin{itemize}
\item $\alpha =1$ implies $\sigma v (S S \rightarrow S^{\dagger} h) \gg \sigma v (S S \Rightarrow XX)$;

\item $\alpha =0.75$ implies $\sigma v (S S \rightarrow S^{\dagger} h) \sim 6 \times \sigma v (S S \Rightarrow XX)$;

\item $\alpha =0.25$ implies $\sigma v (S S \rightarrow S^{\dagger} h) \sim 0.7 \sigma v (S S \Rightarrow XX)$;
\end{itemize}

Interestingly, these scenarios are realized in the $\mathcal{Z_3}$ model. For $\lambda_{SH}=0.1$ and $m_S < 100$~GeV one can obtain the correct relic density while keeping $\alpha > 0.7$. Therefore, the relic density is driven by semi-annihilations. A key aspect of this benchmark model is the dark matter-nucleon scattering is proportional to $\lambda_{SH}^2$. Setting $\lambda_{SH}=0.1$ as mentioned before, one can suppress the spin-independent scattering cross section down to $\sim 10^{-47}$, which is near the expected sensitivity of the XENON1T with two years of exposure. Notice this is not possible in the singlet dark matter model with $\mathcal{Z_2}$ because semi-annihilations are not present and both annihilation and scattering process are grow with $\lambda_{SH}$.

Thus one can sufficiently suppress pair-annihilations along with direct detection limits and possibly explain the GeV excess. A more detailed discussion was carried out in ~\cite{Belanger:2012zr,Cai:2015tam}.

\subsubsection{General $\mathcal{Z}_N$ models}

The very simple setup discussed in the previous subsection provides a viable DM phenomenology and can account for the GeV excess. However semi-annihilation rely on a dimensionfull coupling. In order to have semi-annihilation induced by a dimensionless coupling one should enlarge the spectrum of new states by also introducing an inert scalar $SU(2)$ doublet, so that the scalar potential reads~\cite{Belanger:2012vp,Belanger:2014bga}:

\begin{align}
\label{eq:VZ3}
& V_{\mathcal{Z}_3}=V_0+\frac{\mu^{'}_S}{2}\left(S^3+S^{\dagger\,3}\right)+\frac{\lambda_{S12}}{2}\left(S^2 H_1^{\dagger}H_2+S^{\dagger\,2}H_2^{\dagger}H_1\right)\nonumber\\
& +\frac{\mu_{SH}}{2}\left(S H_2^{\dagger}H_1+S^{\dagger}H_1^{\dagger}H_2\right),
\end{align}where,
\begin{align}
\label{eq:Vgeneral}
& V_0=\mu_1 |H_1|^2+\lambda_1 |H_1|^4+\mu_2 |H_2|^2+\lambda_2 |H_2|^4\nonumber\\
& +\mu_S |S|^2+\lambda_S |S|^4 \nonumber\\
& +\lambda_{S1}|H_1|^2 |S|^2+\lambda_{S2}|H_2|^2 |S|^2\nonumber\\
&+\lambda_{3}|H_1|^2 |H_2|^2+\lambda_4 (H_1^{\dagger}H_2)(H_2^{\dagger}H_1)
\end{align} with $H_1$, $H_2$ and $S$ representing respectively, the SM Higgs doublet, an additional doublet and singlet fields. They are assumed to have charges, respectively, $0,1,1$ under the $\mathcal{Z}_3$ symmetry. 

The term proportional to $\mu_{SH}$ in Eq.\eqref{eq:Vgeneral} will induce a mass mixing between the singlet state $S$ and the neutral component $H_2$ after electroweak spontaneous symmetry breaking. Therefore, our dark matter candidate is a singlet-doublet scalar field under $SU(2)$. 

The presence of a double component is responsible for interactions with gauge bosons which could trigger a too large spin-independent scattering cross section off nuclei due to processes involving the Z boson. Thus one needs to suppress this doublet component. In this model ordinary pair annihilations, $DM DM \rightarrow XX,X=Z,W,h$, are governed by the couplings in Eq.(\ref{eq:Vgeneral}), while semi-annihilations, $DM DM \rightarrow h DM^{*}$, are dictated by the couplings in Eq.\ref{eq:VZ3}). Therefore, one can assume $\mu^{'}_S,\lambda_{S12} \gg \lambda_{S1}$
to render the semi-annihilation the dominant process accounting for the DM relic density and achieve, at the same time, values of the semi-annihilation cross-section in the favored range by the DM excess. 

One can find that for dark matter masses between 120-140 GeV, which is the region of interest for the signals under consideration, keeping $\mu_S^\prime \simeq 200$~GeV and $\lambda_{S1} \simeq 0.01$ we get $\alpha \simeq 1$ as defined in Eq.\eqref{alphaEq} while yielding the correct relic density in agreement with ~\cite{Belanger:2014bga}. Moreover, this same benchmark model leads to a spin-independent dark matter-nucleon scattering of about $10^{-47} cm^{2}$ which is much below current direct detection limits, but within sensitivity of the upcoming XENON-nT experiment \cite{Aprile:2015uzo}.

Viable setups can be obtained by considering also generic $\mathcal{Z}_N, N>3$ groups. For example, by keeping the same field content as the previous case but considering a $\mathcal{Z}_4$ rather than a $\mathcal{Z}_3$ symmetry we can write the potential:
%
%\subsubsection{$Z_4$ symmetric DM}

\begin{align}
\label{eq:VZ4}
& V_{\mathcal{Z}_4}=V_0+\frac{\lambda_S}{2}\left(S^4+S^{\dagger\,4}\right)+\frac{\lambda_5}{2}\left[(H_1^{\dagger} H_2)^2+(H_2^{\dagger} H_1)^2\right]\nonumber\\
&+\frac{\lambda_{S12}}{2}\left(S^2 H_1^{\dagger}H_2+S^{\dagger\,2}H_2^{\dagger}H_1\right)\nonumber\\
&+\frac{\lambda_{S21}}{2}\left(S^2 H_2^{\dagger}H_1+S^{\dagger\,2}H_1^{\dagger}H_2\right)
\end{align}
leading to a similar phenomenology, as the previous case, but with the  difference that there are two distinct DM components, a SM singlet and a doublet.

\subsubsection{Semi-annihilation fermionic DM}

An interaction between three fermions and a Higgs boson is forbidden by Lorentz invariance. As a consequence, fermionic DM can participate to semi-annihilation process only in frameworks in which at least one additional scalar or vectorial DM component is present. A simple option is represent by a model with a scalar $S$ and a fermionic $\psi$ DM candidate~\footnote{A scalar/fermionic multi component DM setup has been also proposed in~\cite{DEramo:2010keq}. In that model the processes $DM DM \rightarrow DM h/Z$ do not take place.}, both singlets under the SM gauge group, described by the following lagrangian~\cite{Cai:2015zza}:
\begin{align} 
&\mathcal{L}=\mathcal{L}_{\rm SM}+\bar \psi \left(i\slashed{\partial}-m_\psi\right)\psi+\partial_\mu S^{\dagger}\partial^\mu S\nonumber\\
&+\frac{1}{2}\left(m_S^2-\lambda_{SH}v^2\right)|S|^2\nonumber\\
& +\left(y S \psi \psi^c+h.c.\right)+\frac{1}{2}\lambda_{SH} H^{\dagger}H |S|^2+\frac{1}{4}\lambda_S |S|^4
\end{align}
where $\psi^c$ is the charge conjugate fermion DM field.
The DM relic density, now a combination of the two components, is determined by two types of semi-annihilation processes, i.e. $\psi S \rightarrow \psi h$ and $\bar \psi \psi \rightarrow h S$, as well as by conversions $\psi \psi \rightarrow SS$ and pair annihilations into SM states (only for the scalar component) $SS \rightarrow {\rm SM\, SM}$. Among these only the processes $S S \rightarrow {\rm SM\, SM}$ and $\psi S \rightarrow \psi h$ have s-wave dominated cross-sections which can then contribute to a signal in gamma-rays. Indeed we find \cite{Cai:2015zza},

\begin{equation}
\sigma v  (S\, S \rightarrow SM\, SM) \sim \frac{\lambda_{S H}^2 }{16 \pi m_S^2}, 
\label{semifermion1}
\end{equation}

\begin{equation}
\sigma v  (\psi\, S \rightarrow \psi\, h) \sim \frac{\lambda_{SH}^2 y^2 v_{SM}^2 (2m_\psi^2+m_\psi m_S)}{ 8 \pi m_S^3 (m_\psi + m_S)^3} ,
\label{semifermion2}
\end{equation}

\begin{equation}
\sigma v  (\psi\, \bar{\psi} \rightarrow S S) \sim \frac{ 3 y^4 }{128 \pi m_\psi^2} \, v^2,
\label{semifermion3}
\end{equation}and,

\begin{equation}
\sigma v (\psi\, \psi \rightarrow \phi h) \sim \frac{\lambda_{S H}^2 y^2 v_{SM}^2}{1024 \pi m_\psi^4} \, v^2 
\label{semifermion4}
\end{equation}where $v_{SM}$ is the SM Higgs vev equal to $246$~GeV.

Therefore, as far as the abundance of $\psi$ is concerned, one can compare Eq.\eqref{semifermion3} with Eq.\eqref{semifermion4} and then realize that by increasing $\lambda_{SH}$ one can make the semi-annihilation process drive the relic density. Although since these processes are velocity suppressed they cannot account for the excesses observed in the galactic center. Although, the singlet scalar can. Notice that the singlet scalar features s-wave annihilations. Notice that on can tune the parameter $y$ to enhance the semi-annihilation processes in Eq.\eqref{semifermion2} to make it dominant. This is important because then on can disconnect the relic density to the spin-independent dark matter-nucleon scattering cross section. For dark matter masses around $100-200$~GeV, XENON1T imposes $\lambda_{SH} < 0.01$ \cite{Queiroz:2014yna}. In the case where only annihilations are present is problematic because this bound imposed by XENON1T on $\lambda_{SH}$ rules out the region of parameter space that reproduces the correct relic density, since they are both governed by the same coupling. Conversely, this semi-annihilation setup separates the relic density from scattering rate and rescues the singlet scalar dark matter, allowing it to have smaller masses in the region of interest for the anti-proton and gamma-ray excess, because one can keep $\lambda_{SH}$ sufficiently small as long as y is large enough.  The GeV excess has not been explicitly investigated in this setup before and we leave for the future a dedicated study.

\subsubsection{Models based on ``dark'' gauge groups}

Vector DM candidates can be easily identified as stable gauge bosons of an additional, with respect to the SM, gauge symmetry. This models have been studied in detail, for example, in~\cite{Gross:2015cwa,Karam:2015jta,Arcadi:2016kmk,Karam:2016rsz,Arcadi:2016qoz}~\footnote{A somehow similar model, based on a``dark SU(3)'' symmetry has been presented also in~\cite{DEramo:2012fou}. In this reference, however, the main ID signal associated to semi-annihilations is represented by gamma-ray lines.}.

The lagrangian associated to the new particle sector can be schematically written as:
\begin{equation}
\mathcal{L}=-\frac{1}{4}F^a_{\mu \nu}F^{a\,\mu \nu}+{\left(D_\mu \phi_i\right)}^{\dagger}\left(D^\mu \phi_i\right)-V(\phi)
\end{equation}
where $F^a_{\mu \nu}$ is the field strength associated to the new gauge bosons ($a$ is the index of the gauge group) $\phi_i$ are the Higgs fields (belonging to a suitable representation of the gauge group) responsible for the spontaneous breaking of the new symmetry, so that all the vector bosons are massive, while $V(\phi)$ is the scalar potential. The scalar potential also contains a  quartic term involving the SM Higgs doublet, of the type $\lambda_{\phi H} |\phi|^2 |H|^2$ which allows, after EW symmetry breaking, for a coupling between the SM and the new particle sector. Under suitable conditions the spontaneous breaking leaves, especially in the case of non Abelian symmetries, residual $\mathcal{Z}_2 \times \mathcal{Z}_2^{'}$ or $\mathcal{Z}_N$ discrete symmetries, hence allowing for semi-annihilation processes involving three DM particles and the SM Higgs boson. For groups larger than $SU(2)$ semi-annihilations involve multiple DM components.

We will discuss in this work the a model based on a dark SU(3) gauge symmetry (a model based SU(2) symmetry would feature an analogous phenomenology and would then represent a sub-case of the one explicitly discussed here).

This model is described by the following scalar potential:
\begin{align}
& \mathcal{V}=\mu_H^2|H|^2
+\mu_1^2|\phi_1|^2
+\mu_2^2|\phi_2|^2
+\left(\mu_{12}^2\phi_1^\dag\phi_2+\mathrm{H.c.}\right)\nonumber\\
& +\frac{\lambda_H}{2}|H|^4
+\frac{\lambda_1}{2}|\phi_1|^4
+\frac{\lambda_2}{2}|\phi_2|^4\nonumber\\
&
+\lambda_{H11}|H|^2|\phi_1|^2
+\lambda_{H22}|H|^2|\phi_2|^2 \nonumber\\
& +\lambda_3|\phi_1|^2|\phi_2|^2
+\lambda_4|\phi_1^\dag\phi_2|^2\nonumber\\
&
+\left(
\lambda_{H12}|H|^2(\phi_1^\dag\phi_2)
+\frac{\lambda_5}{2}(\phi_1^\dag\phi_2)^2\right.\nonumber\\
& \left.+\lambda_{6}|\phi_1|^2(\phi_1^\dag\phi_2)
+\lambda_{7}|\phi_2|^2(\phi_1^\dag\phi_2)
+\mathrm{H.c.}\right),
\end{align}

Where $\phi_1$ and $\phi_2$ are two Higgs fields, belonging to the fundamental representation of SU(3), responsible of the spontaneous breaking of the new gauge symmetry. They cans be written, in the unitary gauge, as:
\begin{equation}
\phi_1=\frac{1}{\sqrt{2}}\left(
\begin{array}{c}
0 \\
0 \\
v_1 + i \varphi_1
\end{array}
\right),\,\,\,
\phi_2=\frac{1}{\sqrt{2}}\left(
\begin{array}{c}
0 \\
v_2+ i \varphi_2 \\
v_3 + i \varphi_3+ v_4 + i \varphi_4
\end{array}
\right)
\end{equation}
where $v_{i=1,2,3,4}$ are the vacuum expectation values of the components of the triplets~\footnote{Notice that in order to achieve a complete breaking of $SU(3)$ the vevs cannot be all equal~\cite{Gross:2015cwa}.}.

As can be easily argued the most general realization of this model features many free parameters and would then result challenging to analyze. A specific, more predictive realization, has been proposed in~\cite{Gross:2015cwa,Arcadi:2016kmk}. It consists in setting $v_3=v_4=0$ and assuming that the SM Higgs doublet $H$ has sizable coupling only with one of the triplets which break $SU(3)$~\cite{Arcadi:2016kmk}. In this setup the eight $SU(3)$ boson present the following mass pattern:
\begin{align}
& m_{A^1}^2=m_{A^2}^2=\frac{1}{4}\tilde{g}^2 v_2^2,\,\,\,m_{A^3}^2=m_{A^4}^2=\frac{1}{4}\tilde{g}^2 v_1^2\nonumber\\
& m_{A^6}^2=m_{A^7}^2=\frac{\tilde{g}^4}{4}(v_1^2+v_2^2) \nonumber\\
& m_{A^3}^2=\frac{\tilde{g}^2}{4}v_2^2\left(1-\frac{\tan\alpha}{\sqrt{3}}\right),\,\,\,m_{A^8}^2=\frac{\tilde{g}^2}{3}v_1^2
\frac{1}{1-\frac{\tan\alpha}{\sqrt{3}}}
\end{align}
where the angle $\alpha$ depends on the two vevs $v_1$ and $v_2$. In this work we will focus on the case $v_2 < v_1$ for which $\sin\alpha \approx \frac{\sqrt{3}}{4}\frac{v_2^2}{v_1^2}$.
The particle spectrum of the new sector is completed by three CP-even scalars, $h_2,h_3,h_4$ (we label $h_1$ the SM-like Higgs) and  a CP-odd scalar state $\chi$~\footnote{One could also consider the case in which CP conservation is not enforced in the dark sector~\cite{Arcadi:2016qoz}}. 

The DM is made by two components: the mass degenerate pair $A^{1\,\mu},A^{2\,\mu}$ and one between $A^{3\,\mu}$ and $\chi$. The latter case has been extensively studied in~\cite{Arcadi:2016kmk}, where it has been found that semi-annihilations are irrelevant for DM phenomenology. We will then here consider the case in which the DM is made by the three vector components $A^{(1,2,3)\,\mu}$.

The relevant phenomenology is, hence, described by the following lagrangian (for simplicity we have omitted mass and kinetic terms):

\begin{align}
& \mathcal{L}=\frac{\tilde{g}m_A}{2}\left(-h_1 \sin\theta+h_2 \cos\theta\right)\nonumber\\
& \times \left(\sum_{a=1,2}A_\mu^a A^{\mu a}+{\left(\cos\alpha-\frac{\sin\alpha}{\sqrt{3}}\right)}^2A_\mu^3 A^{\mu\,3}\right) \nonumber\\
& + \frac{\tilde{g}^2}{8}\left(h_1^2 \sin^2 \theta-2 h_1 h_2 \sin\theta \cos\theta +h_2^2 \cos^2 \theta\right) \nonumber\\
& \times \left(\sum_{a=1,2}A_\mu^a A^{\mu a}+{\left(\cos\alpha-\frac{\sin\alpha}{\sqrt{3}}\right)}^2A_\mu^3 A^{\mu\,3}\right) \nonumber\\
&+ \frac{\left(h_1 \cos\theta+ h_2 \sin\theta\right)}{v_h} \left(2 m_W^2 W_\mu^{-} W^{\mu\,+}+m_Z^2 Z_\mu Z^\mu-\sum_f m_f \bar f f\right)
\end{align}

Where we have defined $m_{A^1}=m_{A^2}=m_A$. $v_h$ is the vev of the SM Higgs.

As already mentioned, we have assumed sizable interactions of the SM Higgs doublet with only one $SU(3)$ multiplet. As a consequence the relevant interactions between the ``dark'' and the ``visible'' sector are mediated only by two scalar mass eigenstates, $h_1$ (mostly SM-like) and $h_2$, whose interactions with the SM states and the DM components are weighted by the mixing angle $\theta$.To further simplify our study we will implicitly assume $m_{h_2}> 2 m_A$. The other CP-even states $h_3$ and $h_4$ are irrelevant for DM phenomenology since their interactions with the SM states are extremely suppressed and, furthermore, are assumed to be decoupled at a higher mass scale.

The DM relic density is determined by solving a system of coupled Boltzmann equations for the two DM components $A^{1\,\mu}=A^{2\,\mu}=A^{\mu}$ and $A^{3\,\mu}$ and results, hence, as the sum of the relative densities of the two components. It is determined by three type of processes: pair annihilations into SM states (for the typical DM values reproducing the GeV excess the main final states are $WW$, $ZZ$ and, far enough from the threshold, $h_1 h_1$~\footnote{As pointed in~\cite{DEramo:2012fou} a high temperature a small contribution is also provided by the $\bar t t$ final state.}), conversion processes $AA \leftrightarrow A^{3} A^{3}$ and semi-annihilation processes $AA \rightarrow A^3 h_1$ and $A A^3 \rightarrow A h_1$.

A good qualitative understanding can be achieved by considering the scaling of these four type of cross sections with the coupling $\tilde{g}$ and the angle $\theta$, as reported in tab.~\ref{tab:darkSU3}~\footnote{We do not consider here the case of possible enhancement, for example through resonances, of the some cross-sections. In the region of parameter space reproducing the GeV excess $m_A,m_{A^3},m_{h_1}$ have comparable values. As a consequence the difference in the cross-sections are mostly determined by the different scaling with $\tilde{g}$ and $\sin\theta$.} (for the $WW$ and $ZZ$ final states we have also reported explicitly the dependence on the $SU(2)$ gauge coupling $g$).

\begin{table}
\begin{center}
\begin{tabular}{|c|c|}
\hline
$WW$, $ZZ$ & $g^2\tilde{g}^2 \sin^2 \theta$ \\
\hline
$hh$ & $\tilde{g}^4 \sin^4 \theta$ \\
\hline
semi-annihilation & $\tilde{g}^4 \sin^2 \theta$ \\
\hline
conversions & $\tilde{g}^4$ \\
\hline
\end{tabular}
\end{center}
\caption{Schematic representation of the parametric dependence on the parameters $\theta$ and $\tilde{g}$ of the different kinds of rates governing the DM relic density in the dark SU(3) model.}
\label{tab:darkSU3}
\end{table}

As evident the different rates depend on different powers of $\sin\theta$ and $\tilde{g}$. The semi-annihilation rate can be made dominant, with respect to pair annihilations, by taking $\tilde{g} \gg g$ and $\sin\theta \ll 1$. We emphasize that this last condition is also favored by the experimental constraint on modification of the SM Higgs sector. We also notice that high values of $\tilde{g}$ would correspond to extremely efficient conversion processes $AA \rightarrow A^3 A^3$, implying that the lightest DM component, $A^3$, retains almost the totality of the DM density fraction. This would imply that the semi-annihilation rate at present times would be negligible, since semi-annihilation processes would require at least one $A^\mu$ in the initial state. To avoid this problem we need so assume $v_2 \ll v_1$ implying $\sin\alpha \simeq 0$ and then that $A^\mu$ and $A^{3\,\mu}$ are mass degenerate so to kinematically suppress conversion processes.

These qualitative arguments have been verified by conducing a numerical analysis through the package MICROMEGAs~\cite{Belanger:2014vza}.

The two DM components feature essentially the same scattering cross-section on nucleons, since $\left(\sin\alpha-\frac{\tan\alpha}{\sqrt{3}}\right) \sim 1$ in our setup. This is of spin-independent type and it is induced by t-channel exchange of the $h_1,h_2$ states. The cross-section reads:

\begin{align}
& \sigma_{Ap}^{\rm SI}=\frac{\mu_{\rm Ap}^2 {\tilde{g}}^2}{4\pi}s_\theta^2 c_\theta^2 {\left(\frac{1}{m_{h_1}^2}-\frac{1}{m_{h_2}^2}\right)}^2 \frac{{\left[f_p Z+f_n (A-Z)\right]}^2}{A^2}\nonumber\\
& \approx 1.8 \times 10^{-47} {\mbox{cm}}^2 \tilde{g}^2 {\left(\frac{\sin\theta}{0.01}\right)}^2
\end{align}
where $\mu_{Ap}=m_p m_A/(m_p+m_A)$ and $f_p \simeq f_n \approx 0.3$. For simplicity we have assumed $m_{h_2} \gg m_{h_1}$ for the numerical estimate.

Similarly to the portal models~\cite{Arcadi:2017kky}, away from s-channel resonances, the DM scattering cross section features the same scaling of the couplings as the annihilation cross-section into SM fermions and gauge bosons. Interestingly the prescription proposed above to increase the importance of the semi-annihilation rate would allow to reduce the impact of Direct Detection constraints.

Unfortunately it has emerged, from our numerical analysis, that, for DM masses within the $2 \sigma$ best fit region of the GeV excess, it possible at most to have $\tilde{g}\simeq 1$ without conflicting with perturbativity bounds on the quartic couplings of the scalar potential. For this value of coupling, accounting also for the requirement of the correct DM relic density, the cross-section corresponding to semi-annihilation processes is at most comparable with the ones corresponding to pair annihilation processes of the two DM components into SM states. Our numerical study suggested, as benchmark assignation with maximal contribution to relic density and Indirect Detection from semi-annihilations, $m_{A} \simeq m_{A^3}=140\,\mbox{GeV}$, $\tilde{g}=1$ and $\sin\theta=0.08$. For this benchmark the two DM components equally contribute to the correct relic density and their pair annihilation as well as their semi-annihilation cross-section have very similar values of around $2 \times 10^{-26}{\mbox{cm}}^3 {\mbox{s}}^{-1}$. The SI DM scattering cross-section on nucleons, of around $10^{-45}\,{\mbox{cm}}^2$, however, exceeds the present limit, as set by the XENON1T experiment~\cite{Aprile:2017iyp}.

It appears hence evident that semi-annihilations cannot contribute, in the dark SU(3) model, to relax the tension with Direct Detection limits in reproduction the gamma-ray excess. We emphasize however that we have considered a specific realization of this scenario and that a conclusive statement requires a more extensive dedicated study.

%Similarly to the models presented in the previous subsection fitting the GeV excess only through semi-annihilation is non-trivial since these processes are accompanied by ordinary pair annihilations as well as, for multi-component setups, by efficient conversion processes. This task is further complicated by the fact that the both pair annihilations, semi-annihilations and conversions are substantially determined by a same coupling, being the gauge coupling of the new sector. A more detailed investigate of this setup will be left for the future.\\

\subsubsection{Accidental $Z_3$ Scalar Dark Matter}

We conclude our theoretical overview by noticing that a similar fit of the GCE, as the one provided by the $DM DM \rightarrow DM h$ semi annihilation processes, could be achieved by considering, as final state, an exotic scalar field $\Phi$, rather than the Higgs boson, provided that the latter dominantly decays into $\bar b b$ final states.

An example of this scenario has been discussed in~\cite{Guo:2015lxa}. Interestingly the discrete $\mathcal{Z}_3$ symmetry, which stabilizes the DM, arises as remnant of a spontaneously broken $U(1)_X$ gauge symmetry (identified with B-L in~\cite{Guo:2015lxa}), by the vev of $\Phi$. The coupling with a complex scalar DM candidate, which has a charge $-2/3$ under $U(1)_X$, responsible for the semi-annihilation processes, is contained in the following scalar potential:
%Another interesting scenario would the generation of a discrete symmetry as result of the spontaneous symmetry breaking of a local symmetry. One could indeed break a $U(1)_X$ symmetry and be left with a remnant $Z_3$ symmetry in the dark sector depending on the $U(1)_X$ charges of the fields. For concreteness, in \cite{Guo:2015lxa} is has been shown that one could break B-L and generate a $Z_3$ symmetry. A complex scalar that has a charge $-2/3$ under $U(1)_X$ becomes stable as a result. The relevant lagrangian in this scenario would encompass,

\begin{eqnarray}
&& V= \lambda_{S \Phi} S^2 \Phi^2 + \lambda_{S H} S^2 H^2 + \lambda_3 \Phi S^3 + \lambda_H | H |^4 \nonumber\\ &&+\lambda_{H\Phi} |H|^2 |\Phi|^2 + \lambda_\Phi |\Phi|^4
\end{eqnarray}
%where $\Phi$ and S are scalar singlets under the SM. 
The crucial term for our reasoning is the one proportional to $\lambda_3$. Similarly to the previously discussed scenarios, one could make the semi-annihilation $S S \Rightarrow S \Phi$ to be the most relevant process for the DM relic density and indirect signal by a suitable assignation of the coupling $\lambda_3$.  
%If $\lambda_3$ is sufficiently large one can boost semi-annihilations. In this setup, the relevant semi-annihilation is $S S \Rightarrow S \Phi$, with $\Phi$ decaying into $b\bar{b}$. 
The coupling of $\Phi$ with the SM fields is originated by the mixing with the SM Higgs. A similar fit of the GCE, as the one discussed in the previous section, could be obtained provided that $2 m_b < m_\Phi < 2 m_W$. Since the mass of this new scalar field do not coincide with the one of the SM Higgs the range of best fit DM masses is different, with respect to the $DM DM \rightarrow DM h$ case, though.

%As we have shown in the previous section from a model independent perspective, this semi-annihilation process provides a good fit to the GeV and antiproton excesses. 
%This conclusion has been also reached in this accidental $Z_3$ scalar dark matter model in \cite{Guo:2015lxa}.

\section{Note on the AMS-02 Anti-Helium Data}

We point out that recently a mild excess observed in anti-Helium \footnote{Check \url{https://indico.cern.ch/event/592392/}.} has been interpreted in terms of WIMPs annihilations \cite{Coogan:2017pwt}. The WIMP properties needed to explain this excess seem different from the ones favored by the anti-proton and gamma-ray excesses. Since the excess is not statistically significant, we have removed it from our reasoning. See \cite{Cirelli:2014qia,Blum:2017qnn} for further discussions in this topic.\\

\section{Conclusions}

In this work we have revisited the excesses observed in gamma-rays and anti-protons toward with Fermi-LAT and AMS-02 satellites.  After briefly discussing the best-fit regions in terms of dark matter annihilations we derived the best-fit regions for the semi-annihilation processes $DM DM \rightarrow DM Z/h$. Pursuing a general approach we have found that the semi-annihilations with one Higgs boson in the final state provide a satisfying fit to the excesses, while still compatible with exclusion limits from Dwarf Spheroidal Galaxies taking into account uncertainties in the dark matter density profile in the galactic center. Similarly to what occurs in the annihilation case, a tension exists with limits from Dwarf Spheroidal Galaxies, thus the need for a large J-factor in the Galactic Center continues. The main advantage of having semi-annihilations as explanation to astrophysical signals and the dark matter relic density is the possibility to break the direct connection to direct detection, allowing one to suppress the dark matter-nucleon scattering cross section without prejudice. In particular, we found, see Fig.3, that the GeV gamma-ray excess favors semi-annihilation cross sections around $3-4 \times 10^{-26} cm^3 s^{-1}$. A similar conclusion applies to the anti-proton excess observed in the AMS-02 data.

We have then briefly reviewed possible theoretical frameworks accounting for the presence of semi-annihilation processes of the type $DM DM \rightarrow DM h$, and discussed under which conditions semi-annihilations can be the dominant contribution to the gamma-ray and anti-proton excesses while yielding the correct relic density and satisfying direct detection constraints. 

\section*{Acknowledgement}
The authors would like to thank Carlos Yaguna, Stefano Profumo, Patrick Draper, Will Shepherd, Christoph Weniger, Fabio Iocco, and Alessandro Cuoco for discussions. FSQ thanks Carlos Pires and Paulo Rodrigues from UFPB, Alvaro Ferraz, Rafael Chaves and Sylvio Quezado from IIF-NATAL for the hospitality in their respective universities where this project was partly carried out. CS is supported by CAPES/PDSE Process 88881.134759/2016-01.

\bibliography{darkmatter}

\end{document}